\documentclass[multphys,vecphys]{svmult}
\usepackage{makeidx}         
\usepackage{graphicx}        
\usepackage{multicol}        

\makeindex             

\begin{document}

\title*{Internal Kinematics of Galaxies: \\3D Spectroscopy on Russian 6m
Telescope.}

\author{A.V. \,Moiseev}

\institute{Special Astrophysical Observatory,    Russian Academy of
Sciences, 369167 Russia \texttt{moisav@sao.ru}}

\titlerunning{Internal Kinematics of Galaxies}
\maketitle

\begin{abstract}
We have considered  some results concerning   gas and stars kinematics of
nearby galaxies recently obtained on the  SAO RAS 6m telescope using the
panoramic spectroscopy methods. The circumnuclear regions of the galaxies were
observed with integral-field spectrograph MPFS. The large-scale ionized gas
kinematics was studied with the  scanning Fabry-Perot interferometer (FPI) in
the multi-mode focal reducer SCORPIO. The main attention is given for
kinematically decoupled regions in the galaxies: bars, spirals, polar disks
and rings.
\end{abstract}

\section{Introduction}

The circumnuclear regions within the first kiloparsec in disk galaxies turn
out to be decoupled on its dynamic characteristics. Using of the technique of
panoramic spectroscopy makes it possible to study in detail the differences in
kinematics of the stellar and gaseous components. I briefly review following
types of kinematically decoupled regions in galactic disks:

\begin{itemize}
\item Non-circular motions caused by the dynamical effects
(spirals, bars, colliding rings).

\item Circular rotation in different planes or directions
     (polar rings, counter-rotating disks).

\item Non-circular motions  caused by a violent starformation (bubbles, high-velocity clouds).
\end{itemize}

\section{Instrumentation}

The study of two-dimensional kinematics of galaxies at  6m telescope of SAO
RAS with  the scanning FPI was started by our colleagues from Marseille
Observatoire (J.~Boulesteix et al.) in cooperation with  team in SAO in the
first half of 1980s. The observations were made with the system CIGALE. Then
the IPCS was replaced by a CCD  and in 2000 the first observations with a new
multimode focal reducer SCORPIO \cite{mois:scorpref} were carried out. Today
FPI mode of SCORPIO provides $6'\times6'$ field of view with a spectral
resolution 1-2.5\AA\, in the H$_\alpha$, [NII], [SII] and [OIII] emission
lines. So, using FPI data  we create monochromatic images as well as fields of
the line-of-sight velocities in these lines (Fig.~\ref{mois:fig1}).

The integral field spectrograph MPFS based on the idea by Victor Afanasiev et
al.: the combination of a lens array with  a bundle of fibers
\cite{mois:afanas90}. First version of the spectrograph was developed in 1990.
Current variant of the MPFS became operational at the 6 m telescope since 1998
\cite{mois:mpfs}. Fibers transmit light from $16\times16$ square elements of
the galaxy image to the slit of the spectrograph, together with 16 additional
fibers that transmit the sky background  taken away from the galaxy. The
angular scale is $1''$ and the spectral resolution is $4-12$\AA.  Using MPFS
for kinematics of galaxies allows us to build velocity fields of ionized gas
as well as of their stellar component.

The  descriptions  of our spectrographs are  available at SAO RAS web page: 
\verb"http://www.sao.ru/hq/lsfvo/"

\begin{figure}
\centering
\includegraphics[width=11 cm]{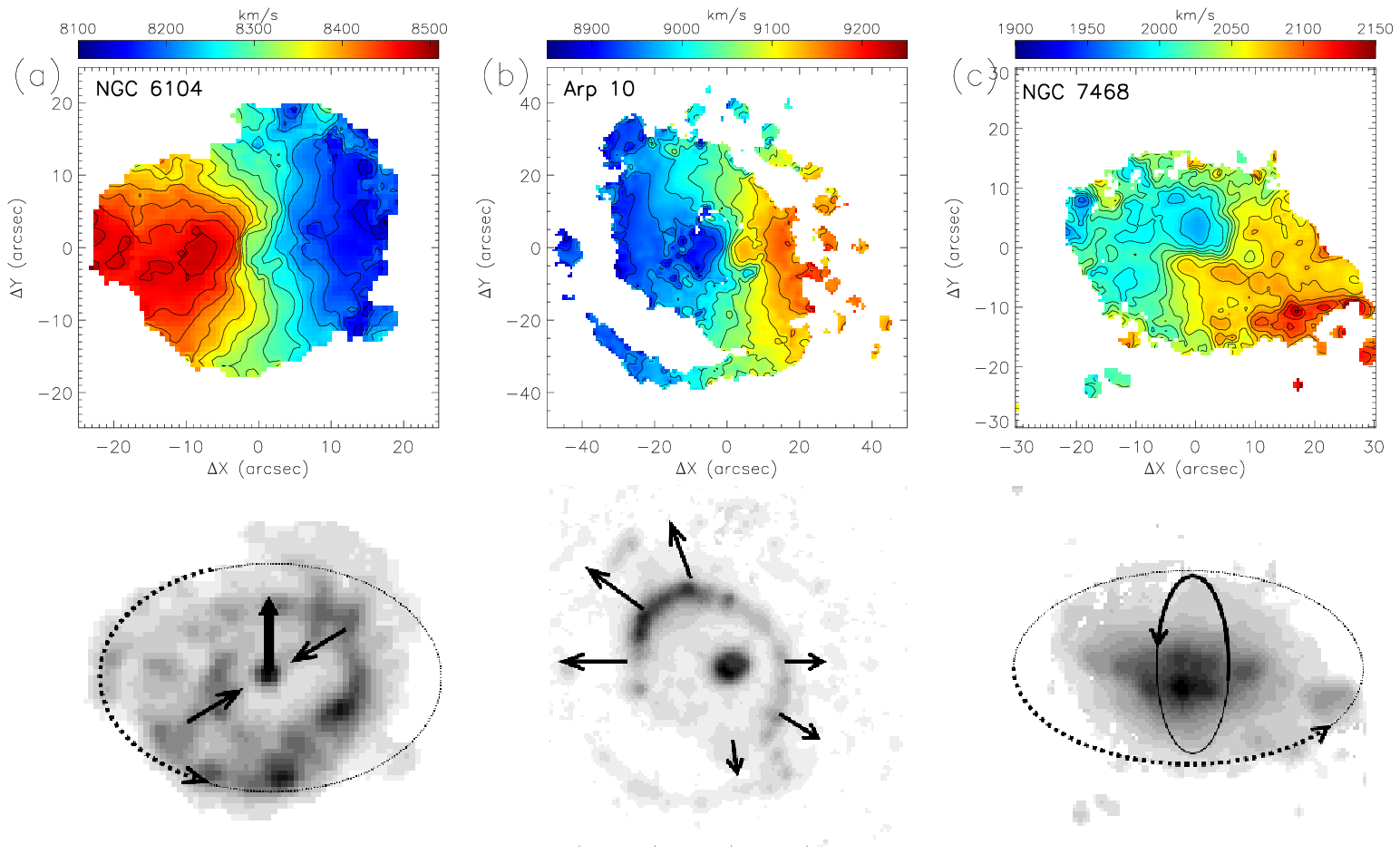}
\caption{ FPI velocity fields of ionizing gas  and H$_\alpha$ images with the
sketch of gas motions: the radial inflow along a bar and the nuclear jet
outflow in Sy NGC~6104 (a); the expansion of the emission ring in Arp~10 (b);
polar disk in NGC~7468 (c)} \label{mois:fig1}
\end{figure}

\section{The Examples of the Objects}

\subsection{Barred and Double-barred Galaxies}

The values of non-circular motions in barred galaxies can be estimate in the
detailed analysis of their velocity fields. For instance, Sy galaxy NGC~6104
where the ionized gas radial inflow motions along the bar co-exist with nuclear
outflow caused by  jet-clouds interactions, \cite{mois:smir}
 (Fig.~\ref{mois:fig1}a).

Recently we have observed a sample of candidate double-barred galaxies and
suggest that these objects  are, in fact, galaxies with very different
circumnuclear structure \cite{mois:mois04}. The majority of the observed
morphological and kinematic features in the sample may be explained without
the secondary bar hypothesis. Three cases of inner polar disks, one
counter-rotating gaseous disk and seven nuclear disks  nested in large-scale
bars were found in this work.

\subsection{Counter-rotation}

\begin{figure}
\centering
\includegraphics[width=11 cm]{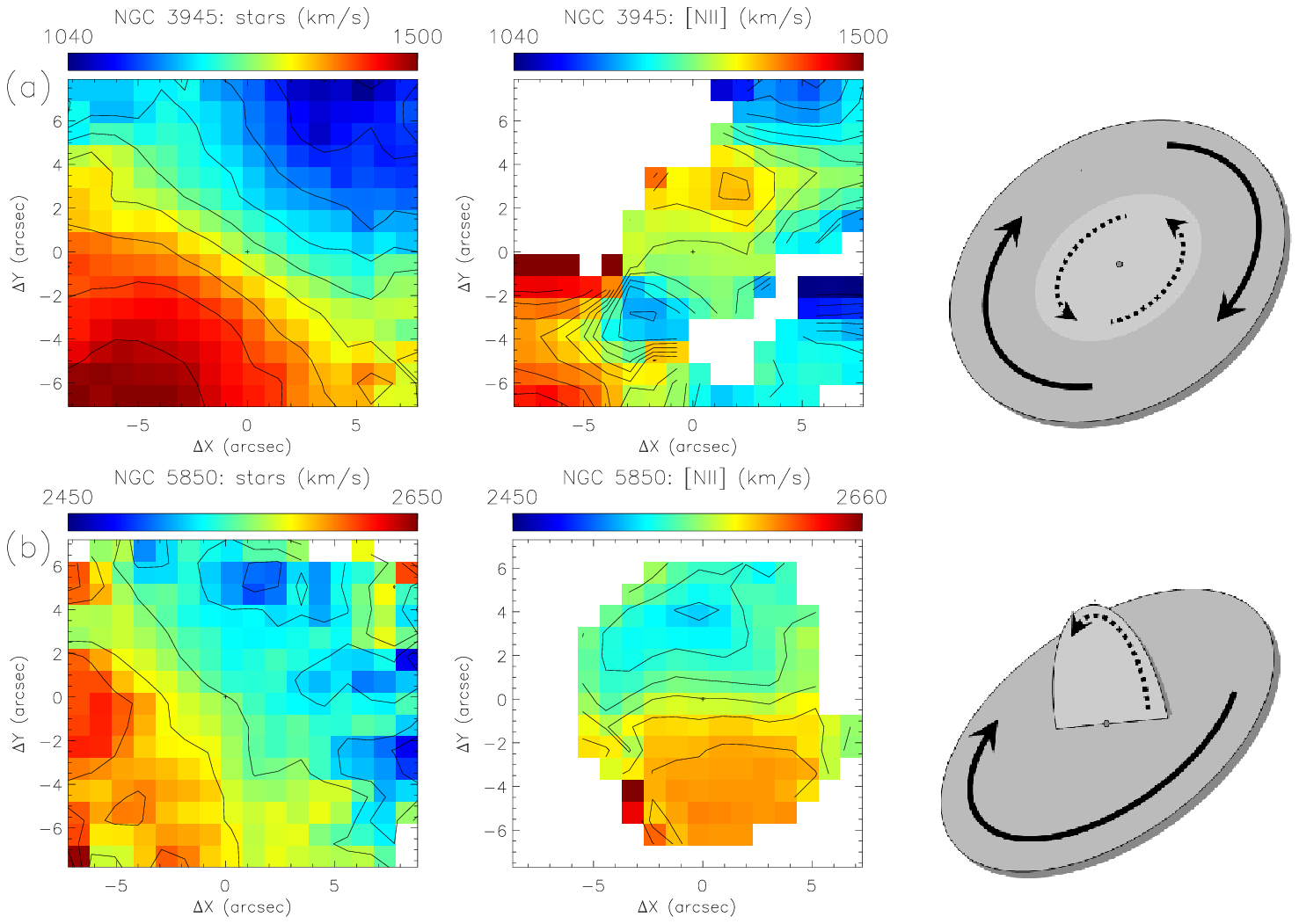}
 \caption{MPFS data (from \cite{mois:mois04}): the velocity fields of stars and ionized gas for the circumnuclear
region of NGC~3945 (a, counter-rotated disk) and NGC~5850 (b, inner polar
disk). The right panels show the sketch of motions of gas and stars.}
\label{mois:fig2}       
\end{figure}

The type of the motion of gas clouds  may noticeably differ from the rotation
of the stellar component even in `quiet' galaxies without  AGN. The
characteristic case  is a lenticular galaxy NGC~3945 (Fig.~\ref{mois:fig2}a).
The velocity field of the stars in the circumnuclear disk  shows the normal
circular rotation. For the gas velocity field the situation is more
complicated. In $r<6''$ (0.5 kpc), the line-of-sight velocities of ionized gas
are  in the maximum amplitude to those of stars but opposite in sign.  On
larger distances the direction of rotation of gas changes abruptly and
coincides with the stellar component rotation. The large-scale FPI velocity
field confirms the fact of normal gas rotation at large radii,  up to 11 kpc.
Therefore, we have detected a circumnuclear disk of ionized gas rotating in
the opposite direction with respect to the stellar component,
\cite{mois:mois04}. This  is probably attributable to a merger of an accreted
gaseous cloud with the corresponding direction of an angular momentum
\cite{mois:bertola}.

\subsection{Collisional Rings}

Collisional ring galaxies represent a class of objects in which nearly
circular density waves are driven into a disk as a result of an almost face-on
collision with another galaxy. We have observed with FPI a pecular galaxy
Arp~10, which has two rings (the inner and outer one), and extended outer
arc.  The H$_\alpha$ velocity field  shows evidence for significant radial
motions in both outer and inner galactic rings. We fit a model velocity field
taking into account the regular rotation and projection effects.  The
expansion velocity of the NW part of the outer ring exceeds 100\,km/s, whereas
it attain only 30\,km/s at the SE part \cite{mois:mois05}. Therefore, the
asymmetric shape of the outer ring (fig.~\ref{mois:fig1}b) may be caused by a
systematic difference in the ring expansion velocity and collisional origin of
the rings is a proven fact.

\subsection{Polar Rings and Nuclear Polar Disks.}

Recently the team in St.Petersburg university (V. Hagen-Thorn, L. Shaliapina,
V. Yakovleva) in a collaboration with us attempt to observe with FPI a gas
kinematics in candidate polar-ring galaxies. The interesting results were
already obtained. For example, they detected an inner gaseous disk  whose
rotation plane is almost perpendicular to the plane of the `main' galactic
disk in   NGC~7648 \cite{mois:shal}. The Fig.~\ref{mois:fig1}c shows the sharp
turn of isovelocities in the galactic circumnuclear region. The central
collision of NGC~7468 with a gas-rich dwarf galaxy and their subsequent
merging seem to be responsible for the formation of the disk.

In the barred galaxy NGC~5850 the direction of rotation measured from the
stellar component coincides with the line of nodes of the disk whereas in the
ionized gas it differs by more than $50^\circ$ and coincides with the position
angle of the central isophotes. Such a behavior is typical for a disk inclined
to the galactic plane. A more reasonable assumption is that the gas, at $r <1$
kpc, rotates in a polar plane with respect to the global galactic disk
\cite{mois:mois04}. In this case, the polar gaseous disk lies  orthogonal to
the major axis of the bar. It is remarkable that similar polar mini-disks
inside the large-scale bars or the three-axial bulge have already been
detected in about twenty galaxies (see \cite{mois:sil04} and the contribution
by Olga Sil'chenko in this volume).

\begin{figure}
\centering
\includegraphics[width=11 cm]{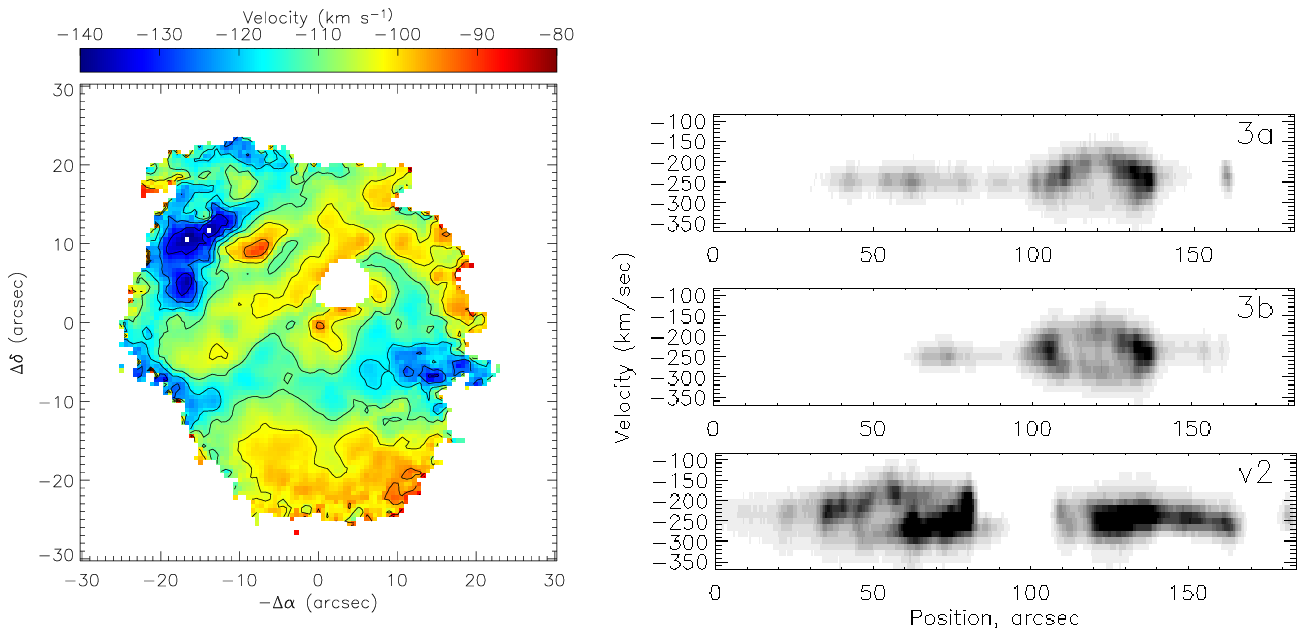}
\caption{Gas kinematics in dwarf  galaxies: (left) -- velocity field of
VIIZw403 \cite{mois:loz06}, (right) -- the position velocity diagrams trough
starforming regions in IC~1613. The remarkable velocity ellipsoids correspond
to expansion of H$_\alpha$ envelopes.}
\label{mois:fig3}       
\end{figure}

\subsection{Starformation in Dwarf Galaxies.}

A burst of starformation also produces  non-circular gas motions triggered by
the combined effect of stellar winds and supernova explosions in rich stellar
groupings. In dwarf galaxies the formation of giant multi-shell complexes
around stellar groupings can proceed unhindered. See, for example,
observations with SCORPIO-FPI of the nearby irregular galaxy IC1613
\cite{mois:loz03}: the authors refined the expansion velocities  of individual
shells of the ionized and neutral gas (Fig.~\ref{mois:fig3}). In such galaxies
the main part of line-of-sight velocities connects with an expansion of  HII
regions, frequently without any rotation (like VIIZw403, see also
Fig.~\ref{mois:fig3}, left)

I'll like to thank V. Afanasiev, S. Dodonov,  O. Sil'chenko and  A. Smirnova
for their help and numerous discussions.  This work was partially supported by
the RFBR grant No.~05-02-16454. I also thank the Russian Science Support
Foundation and the Organizing Committee of the workshop.

 \printindex
\end{document}